\documentclass[pre,twocolumn,showpacs,preprintnumbers,amsmath,amssymb,letter]{revtex4}

\usepackage{graphicx}
\usepackage{dcolumn}
\usepackage{bm}
\usepackage{subfigure}

\begin{document}


\title{Steady state velocity distributions of an oscillated granular gas}
\author{Sung Joon Moon}
\email{moon@chaos.utexas.edu}
\author{J. B. Swift}
\email{swift@chaos.utexas.edu}
\author{Harry L. Swinney}
\email{swinney@chaos.utexas.edu}
\affiliation{Center for Nonlinear Dynamics and Department of Physics,
          University of Texas, Austin, Texas 78712}
\date{\today}

\begin{abstract}
We use a three-dimensional molecular dynamics simulation to study
the single particle distribution function of a dilute granular gas
driven by a vertically oscillating plate at high accelerations
($15g - 90g$).
We find that the density and the temperature fields are essentially
time-invariant above a height of about 35 particle diameters,
where typically 20\% of the grains are contained.
These grains form the nonequilibrium steady state granular gas with
a Knudsen number unity or greater.
In the steady state region, the distribution function of horizontal
velocities (scaled by the local horizontal temperature) is found
to be nearly independent of height, even though the hydrodynamic
fields vary with height.
The high energy tails of the distribution functions are described by
a stretched exponential $\sim \exp(-{\cal B}c_x^{\alpha})$,
where $\alpha$ depends on the normal coefficient of restitution $e$
($1.2 < \alpha < 1.6$), but $\alpha$ does not vary for a wide range of
the friction parameter.
We find that the distribution function of a {\it frictionless} inelastic
hard sphere model can be made similar to that of a frictional model by
adjusting $e$. However, there is no single value of $e$ that
mimics the frictional model over a range of heights.

\end{abstract}


\maketitle

\nobreak

\section{introduction}

A dilute gas in thermal equilibrium is sufficiently characterized
by the pressure and temperature and is described by a simple relation,
the equation of state.
However, when a gas is far from equilibrium, there is no general,
finite set of variables specifying the state.
The single particle distribution function $f({\bf r},{\bf v},t)$ is
often sufficient to characterize the statistical properties of a dilute
nonequilibrium gas when correlations are negligible.
Given this function, other quantities, such as moments
of the distribution and transport coefficients, can be evaluated.
Dilute granular materials subject to an external forcing exhibit
gaseous behaviors that share many analogies with a molecular gas,
and they are often called granular gases.
Such a granular gas is always far from equilibrium due to the dissipative
collisions, and the deviation of $f({\bf r},{\bf v},t)$ from the
Maxwell-Boltzmann (MB) distribution has been of great interest in
recent years~\cite{warr95,olafsen98,losert99,kudrolli00,rouyer00}.

The velocity distributions of a vibro-fluidized granular gas were
first measured by Warr {\it et al.}~\cite{warr95}; they studied
the distribution functions of grains confined between two transparent
plates and concluded that the distribution was consistent with
the MB distribution function. Recently, the same system has been
studied by Rouyer {\it et al.}~\cite{rouyer00}, who found a
{\it universal} distribution function of the form
$\sim \exp(-B|{\bf v}|^{1.5})$ for the {\it entire}
range of velocities studied, where $B$ was a parameter.
The authors reported that this functional form fit their
measurements for a wide range of oscillation parameters for
various materials; thus the granular temperature,
the second moment of the distribution, was the only parameter
of the distribution function.

There have been numerical studies of vibrated inelastic hard disks,
subject to a saw-tooth type oscillation, in the presence of
gravity~\cite{brey03} and in the absence of gravity~\cite{barrat02}.
Such forcing is often used in theoretical studies as a simplification
of the sinusoidal oscillation used in experiments, assuming that
the asymptotic behavior of the hydrodynamic fields far from the plate
is the same; however, it is not known {\it a priori} how far from the
oscillating plate one must be in order for this assumption to be valid.

In this paper, we perform a simulation that is as close as possible to
three-dimensional experiments on vertically oscillated granular
gases. We use a previously validated molecular dynamics (MD)
simulation~\cite{bizon98,moon02a}. The hydrodynamic fields are
oscillatory near the plate, and their oscillatory behavior decays with
height. Above some height, the fields are not correlated with the
oscillation of the plate and are essentially time-invariant.  We study
the distribution functions in this nonequilibrium steady state region.
To focus on the distributions due to the intrinsic dynamics of the
granular gas, we do not impose sidewalls or include air.  We also
study how the distribution changes with the friction.  In many
theoretical or numerical studies of granular fluids, granular
materials are modeled as frictionless inelastic hard disks or spheres;
however, no granular materials are frictionless, in the same way that
none of them are elastic.  We check if the role of friction can be
incorporated into the inelasticity by adjusting the value of the
normal coefficient of restitution.  In this paper we discuss the
distributions only in the steady state region; those in the
oscillatory region near the plate will be discussed in a separate
paper~\cite{moon03vdf2}.

The rest of the paper is organized as follows.
In Section II, the system under consideration, the data analysis
method, and the collision model are described.
Results are presented in Section III and discussed in Section IV.

\section{method}

\subsection{System and data analysis}

We use both a frictional and frictionless, inelastic hard sphere
MD simulation. We consider 133 328 monodisperse spheres of unit
mass and of diameter $\sigma = 165~\mu$m in a container with square
bottom of area 200$\sigma\times 200\sigma$ (the average depth of
the layer at rest is approximately 3$\sigma$), where periodic
boundary conditions are imposed in both horizontal directions.
We choose the same particle size as in Ref.~\cite{bizon98},
as the patterns were quantitatively reproduced for a wide range of
oscillation parameters with this particle size;
however, as long as the collision model is valid, all the length
scales can be normalized by $\sigma$.
We assume the bottom plate of the container is made of the same
material as grains; we use the same material coefficients for the
inelasticity and the friction as grains.
The bottom plate is subject to a vertical sinusoidal oscillation
with an amplitude $A$ and a frequency $f$.
We vary the oscillation parameters in the range of
$3\sigma < A < 10\sigma$ and $40~{\rm Hz} < f < 170~{\rm Hz}$,
which approximately corresponds to
$0.35~{\rm m/s} < V_{max}~(= 2\pi fA) < 0.75~{\rm m/s}$
and $15g < a_{max}~[= A(2\pi f)^2] < 90g$,
where $g$ is the acceleration due to gravity.
We check that with our parameters no mean flow develops and that
grains rarely reach to the top, which is fixed at 300$\sigma$.

Hydrodynamic fields and the distribution functions are analyzed
by binning the box into horizontal slabs of height $\sigma$,
as the system is invariant under the translation in both horizontal
directions, in the absence of any mean flow.
We use the granular volume fraction $\nu$ for the density,
which is the ratio of the volume occupied by grains to the volume
of each horizontal slab.
We consider the following three granular temperatures separately:

\begin{eqnarray}
T_x &=& {1 \over 2} \left<\left(v_x-\left<v_x\right>\right)^2+\left(v_y-\left<v_y\right>\right)^2\right>,\\
T_z &=& \left<(v_z-\left<v_z\right>)^2\right>,\\
T &=& {1 \over 3} \left<|{\bf v}-\left<{\bf v}\right>|^2\right>\\
&=& {1 \over 3} \left(2T_x+T_z\right),
\end{eqnarray}
where $x$ and $y$ are horizontal directions that are indistinguishable,
$z$ is the vertical direction, {\bf v} is a velocity vector for each grain,
and the ensemble average $\left<~\right>$ is taken over the particles
in the same bin at the same phase angle during 40 cycles,
after initial transients have decayed.
We define the scaled horizontal velocity to be
\begin{equation}
\label{scale}
c_x = (v_x - \left< v_x \right>)/\sqrt{2T_x}.
\end{equation}

\subsection{collision model}

We implement the collision model that was originally proposed by
Maw {\it et al.}~\cite{maw}, simplified by Walton~\cite{walton}, and
then experimentally tested by Foerster {\it et al.}~\cite{foerster}.
This model updates the velocity after a collision according to
the three parameters,
the normal coefficient of restitution $e$ ($\in [0,1]$),
the coefficient of friction $\mu$, which relates the tangential
force to the normal force at collision using Coulomb's law
and then determines the tangential coefficient of restitution $\beta$
($\in [-1,1]$), and
the maximum tangential coefficient of restitution $\beta_0$,
which represents the tangential restitution of the surface velocity
when the colliding particles slide discontinuously at the contact point.

At collision, it is convenient to decompose the relative colliding
velocities into the components normal (${\bf v}_n$) and tangential
(${\bf v}_t$) to the normalized relative displacement vector
$\hat{\bf r}_{12} \equiv ({\bf r}_1 - {\bf r}_2)/|{\bf r}_1 - {\bf r}_2|$,
where ${\bf r}_1$ and ${\bf r}_2$ are displacement vectors of grains
1 and 2, and the same notation is used for ${\bf v}$:
\begin{eqnarray}
{\bf v}_n &=& ({\bf v}_{12}\cdot \hat{\bf r}_{12})\hat{\bf r}_{12}
\equiv v_n\hat{\bf r}_{12},\\
{\bf v}_t &=& \hat{\bf r}_{12}\times ({\bf v}_{12}\times \hat{\bf r}_{12}) = {\bf v}_{12} - {\bf v}_n.
\end{eqnarray}
The relative surface velocity at collision, ${\bf v}_s$,
for monodisperse spheres of diameter $\sigma$ is
\begin{equation}
{\bf v}_s = {\bf v}_t + {\sigma \over 2}\hat{\bf r}_{12} \times
({\bf w}_1+{\bf w}_2) \equiv v_s\hat{\bf v}_s,
\end{equation}
where ${\bf w}_1$ and ${\bf w}_2$ are the angular velocities
of grains 1 and 2, respectively.

For monodisperse spheres of diameter $\sigma$ and unit mass,
the linear and angular momenta conservations and the definitions of
the normal coefficient of restitution $e \equiv -v_n^*/v_n$ and the
tangential coefficient of restitution $\beta \equiv -v_s^*/v_s$
(post-collisional velocities are indicated by superscript $*$,
and pre-collisional values have no superscript) give the changes
in the velocities at the collision:
\begin{eqnarray}
\Delta {\bf v}_{1n} &=& -\Delta  {\bf v}_{2n} = {1\over 2}\left(1+e\right){\bf v}_n,\\
\Delta {\bf v}_{1t} &=& -\Delta  {\bf v}_{2t} = {K\left(1+\beta\right)\over 2\left(K+1\right)}{\bf v}_s,\\
\Delta {\bf w}_1 &=& -\Delta  {\bf w}_2 = {\left(1+\beta\right)\over \sigma\left(K+1\right)}\hat{\bf r}_{12}\times{\bf v}_s,
\end{eqnarray}
where $K = 4I/\sigma^2$ is a geometrical factor relating
the momentum transfer from the translational degrees of freedom
to rotational degrees of freedom, and $I$ is the moment of inertia
about the center of the grain.
For a uniform density sphere, $K$ is $2/5$.

We use a velocity-dependent normal coefficient of restitution
as in Ref.~\cite{bizon98}, to account for the viscoelasticity
of the real grains:
\begin{equation}
e = {\rm max}\left[e_0,1 - (1-e_0)\left({v_n\over \sqrt{g\sigma}}\right)^{3/4}\right],
\end{equation}
where $e_0$ is a positive constant less than unity.
Since we 
impose high forcing ($V_{max} > 0.35$ m/s while
$\sqrt{g\sigma} = 0.04$ m/s)
the collision probability for relative colliding velocities $v_n$
less than $\sqrt{g\sigma}$ is small, and using a velocity-independent
$e$ does not result in any noticeable difference,
compared to using a velocity-independent one $e = e_0$;
the same was true in Ref.~\cite{bougie02}.
We use the symbol $e$ for $e_0$ hereafter.

In collisions of real granular materials, not only is the relative
surface velocity reduced, but also the stored tangential strain energy
in the contact region can often reverse the direction of the relative 
surface velocity.
To account for this effect, the tangential coefficient of restitution
$\beta$ could be positive, leading to the range of $\beta$ as $[-1,1]$.
There are two kinds of frictional interaction at collisions, sliding
and rolling friction, which are accounted for by the following formula
for $\beta$:
\begin{equation}
\beta = {\rm min}\left[\beta_0,-1 + \mu\left(1+e\right)\left(1+{1\over K}\right){v_n\over v_s}\right],
\end{equation}
where $\beta_0$ is the maximum tangential coefficient of restitution.
For sliding friction, the tangential impulse is assumed to be
the normal impulse multiplied by $\mu$.
When $\beta$ is identically negative unity (or simply $\mu = 0$),
this model reduces to the frictionless interaction.
For the special case $v_s = 0$, the collision is treated as frictionless.
This friction model is still a simplification of the real frictional
interaction; there is no clear-cut distinction between the two types
of frictions for real grains, and even a transfer of energy from
the rotational to translational degrees of freedom, which results in
$e$ larger than unity, has been observed~\cite{louge}.
However, this collision model is accurate enough to reproduce many
phenomena, including standing wave pattern formation in vertically
oscillated granular layers, when the parameters are properly chosen.
In Refs.~\cite{bizon98} and \cite{moon02a}, $e = 0.7,~\beta_0 = 0.35$,
and $\mu = 0.5$ were used.

\begin{figure}[t!h]
{\includegraphics[width=0.805\columnwidth]{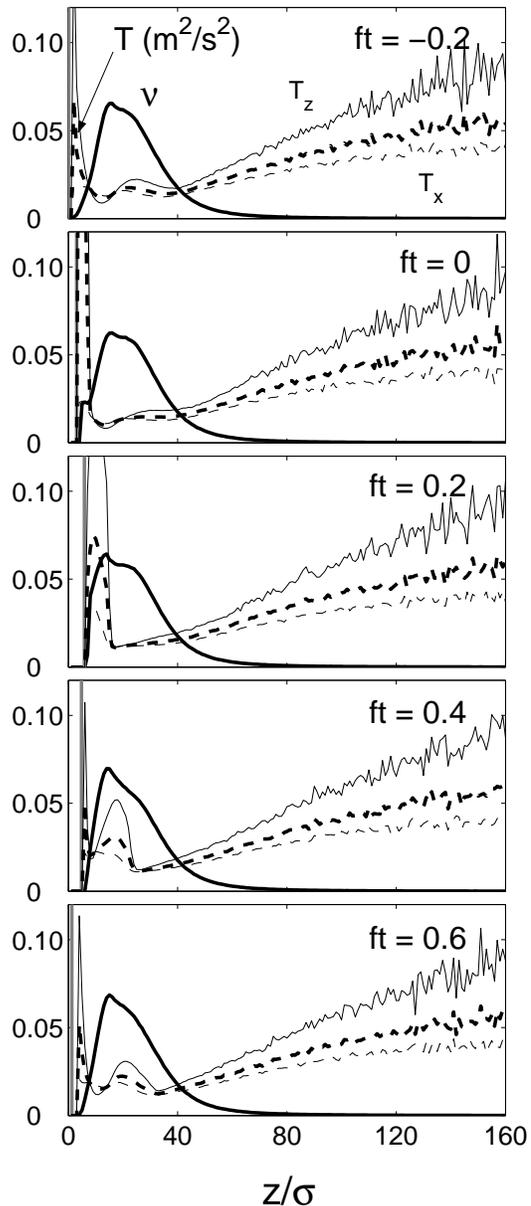}}
\caption{
\label{FF}
The volume fraction $\nu$ (thick solid line) and the granular
temperature $T$ (thick dashed line) as a function of height
at different times during
a cycle, where $e = 0.9,~\beta_0 = 0.35,~\mu = 0.5$,
$V_{max} = 0.55$ m/s ($a_{max} = 60g$, $A = 3\sigma$, and $f = 169$ Hz),
and $ft$ is set to zero when the plate is
at the equilibrium position moving upward.
Above some height ($z/\sigma \approx 35$),
the hydrodynamic fields do not vary much in time.
The horizontal temperature $T_x$ (thin dashed line) is smaller
than the vertical temperature $T_z$ (thin solid line).
The container bottom is indicated by the vertical gray line.
}
\end{figure}

\begin{figure}[h]
{\includegraphics[width=0.8\columnwidth]{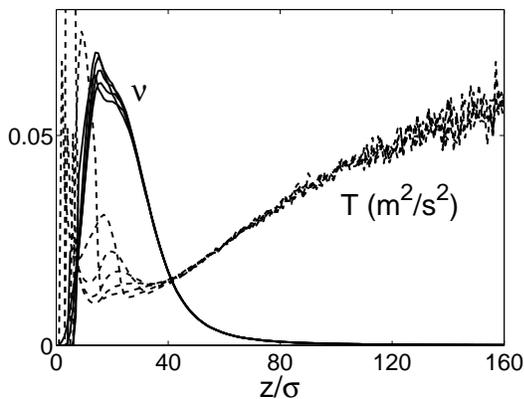}}
\caption{
\label{T-INV}
This superposition of the volume fraction and temperature fields
at five different times in a cycle (see Fig.~\ref{FF}) illustrates
that the fields are nearly time-independent above $z/\sigma \approx 35$,
which we call the steady state region.
}
\end{figure}

\begin{figure}[b]
{\includegraphics[width=0.8\columnwidth]{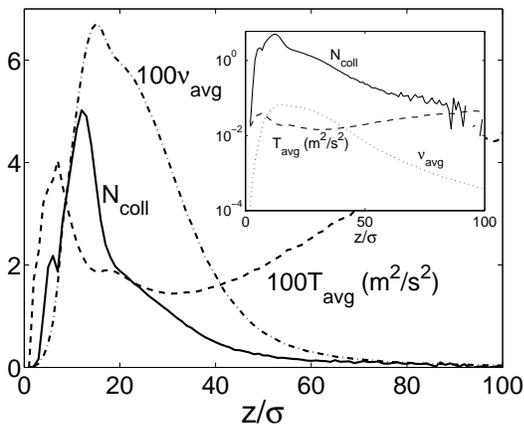}}
\caption{
\label{CPP}
The number of grain-grain collisions per grain during
a cycle $N_{coll}$ (solid line), and time-averaged volume fraction
$\nu_{avg}$ (dot-dashed line, multiplied by 100)
and time-averaged granular temperature $T_{avg}$
(dashed line, multiplied by 100), over sixty different equally spaced
times during a cycle.
$N_{coll}$ is less than 1 in the steady state region ($z/\sigma > 35$).
Inset :
The same quantities (without multiplications) on a logarithmic scale.
}
\end{figure}

\section{results}
\subsection{Hydrodynamic fields and steady state}
Due to the oscillatory boundary forcing, the hydrodynamic fields,
the volume fraction $\nu$ and the granular temperatures ($T,~T_x$,
and $T_z$), depend on height $z$ and time $t$ (Fig.~\ref{FF})
near the oscillating plate; the temperatures exhibit stronger
oscillatory behaviors than the density.
Since the energy is injected mainly through the vertical
velocities, the granular temperature is anisotropic,
as illustrated in Fig.~\ref{FF}; $T_z$ is larger than its horizontal
counterpart $T_x$, and the former is significantly larger near
the bottom plate, where the hydrodynamic fields are oscillatory.
The vertical temperature increases almost linearly with height for
$z/\sigma > 35$; however, the temperature $T$ increases slower than
linearly, as the slope of $T_x$ decreases with height and $T_x$
 levels off for $z/\sigma > 120$ (Fig.~\ref{FF}).
A similar increase of the temperature with height was observed
in an open system of frictionless inelastic hard disks or spheres
subject to a thermal bottom heating~\cite{soto99} and a saw-tooth
type vibration~\cite{brey01}.
We characterize the oscillation parameters only by $V_{max}$,
as we observe for the parameters in our study that the hydrodynamic
fields in the steady state are nearly the same for the same $V_{max}$,
even for different combinations of $a_{max}$ and $f$; such scaling
behavior was also observed in Ref.~\cite{lee95}.

During each cycle, a normal shock forms at the impact from
the bottom plate and propagates upward~\cite{bougie02}.
As the shock propagates, it decays and becomes undetectable above
some height ($z/\sigma \approx 35$), rather than propagating up
through the entire granular media
(which was the case in Ref.~\cite{bougie02}).
Above this height, the hydrodynamic fields are invariant in time,
and the granular gas forms a nonequilibrium steady state
(Fig.~\ref{T-INV}), where about 20\% of the grains are contained
in this case; this fraction depends on the oscillation parameters.

With the parameters used in this paper the granular temperatures
are nonzero throughout the cycle, as grains do not solidify after
the shock passes through, in contrast to the case in Ref.~\cite{bougie02}.
As a result, when the bottom plate moves down, the granular gas
expands, and an expansion wave propagates downward (see the
temperature peaks near the plate for $ft > 0.4$ in Fig.~\ref{FF}).

\begin{figure}[t]
{\includegraphics[width=0.75\columnwidth]{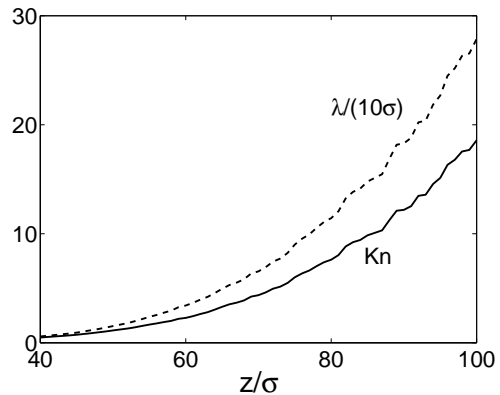}}
\caption{
\label{mfpKn}
The mean free path $\lambda$, estimated from a formula
for a gas of hard spheres, and the Knudsen number $Kn$,
estimated by using $\lambda$ and the length scale of the density
$l_{\nu}$ (see text), in the steady state region.}
\end{figure}

We count the number of grain-grain collisions per grain
during a cycle ($N_{coll}$ in Fig.~\ref{CPP}), and
find that $N_{coll}$ is less than unity in the steady state
region; the granular gas in the steady state is nearly collisionless.
We estimate the mean free path using the formula for a gas of hard
spheres $\lambda(z)/\sigma = (2\sqrt{2\pi}n\sigma^3)^{-1} =
\sqrt{\pi}/(12\sqrt{2}\nu)$ (where $n$ is the number density)~\cite{huang},
which ranges between 5 and 280 for $40 < z/\sigma < 100$ (Fig.~\ref{mfpKn}).
When we estimate the mean free path using the measured collision
frequency and the thermal speed, we get a similar result.
We fit the density with piecewise exponential functions
[$\sim \exp(-(z-z_0)/l_{\nu})$] in the steady state region,
and obtain a hydrodynamic length scale $l_{\nu}/\sigma$ between
12 and 15 for $40 < z/\sigma < 100$.
We obtain a similar length scale from piecewise linear fitting
of the temperature $T$ in the same region.
We calculate the Knudsen number $Kn$, defined as the ratio
of the mean free path to the length scale of the macroscopic
gradients~\cite{kogan69}, using $\lambda(z)$ and $l_{\nu}$;
$Kn$ ranges from 0.5 to 20 in the region $40 < z/\sigma < 100$
(Fig.~\ref{mfpKn}).

\subsection{Height-independence of the distribution}
The distribution of scaled horizontal velocities should be symmetric
as a consequence of the symmetry of the system.
We calculate the skewness of the distribution,
$\gamma_3 = {\cal M}_3/{\cal M}_2^{3/2}$,
where ${\cal M}_n$ is the $n^{th}$ moment of the distribution

\begin{figure}[t]
{\includegraphics[width=0.8\columnwidth]{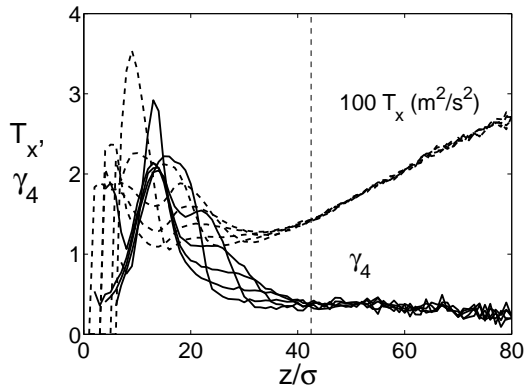}}
\caption{
\label{TK}
Above some height ($z/\sigma \approx 42$, indicated by a vertical
dashed line), the kurtosis $\gamma_4$ is nearly time-invariant.
The horizontal granular temperature $T_x$ (dashed lines) and $\gamma_4$
(solid lines) of the horizontal velocity distribution function are
shown at five different times during a cycle ({\it cf.} Fig.~\ref{FF}).
}
\end{figure}

\begin{figure}[b]
{\includegraphics[width=0.8\columnwidth]{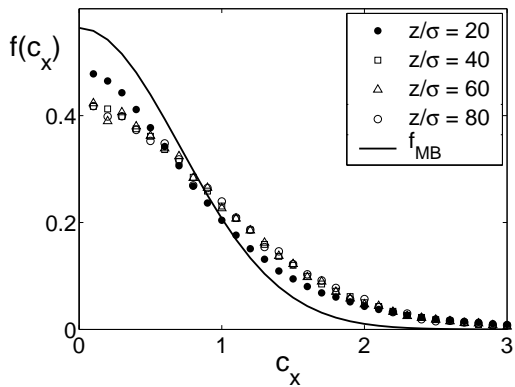}}
\caption{
\label{M4}
The horizontal velocity distribution functions at four
different heights (compared with $f_{MB}$, the solid line)
obtained at $ft = -0.2$.
There is no noticeable difference among the distributions
in the steady state region, $40 < z/\sigma < 80$.
}
\end{figure}

\begin{equation} 
\label{EQmoment}
{\cal M}_n = \int c_x^nf(c_x) dc_x,
\end{equation} 
and check that $|\gamma_3|$ is less than 0.01
for all the distributions we study.
The lowest order deviation from the MB distribution is characterized
by the flatness of the distribution, which is called the fourth
cumulant or the kurtosis. It was used to quantify the deviation from
the MB distribution of the homogeneously cooling
state~\cite{vannoije98,brilliantov00,nie02},
the homogeneously heated state~\cite{vannoije98,moon01},
and granular gases subject to a boundary forcing~\cite{brey03,blair01}.
The kurtosis $\gamma_4$ ($\equiv {\cal M}_4/{\cal M}_2^2 -3$) is
defined so that it vanishes for the MB distribution, and
we find that it also does not change in time above some
height (Fig.~\ref{TK}). 
Further, in the steady state region, $\gamma_4$ is nearly
independent of the height, even though both the density and
the temperature change; the distributions at different heights in
the steady state region are hardly distinguishable (Fig.~\ref{M4}).
Also, the kurtosis in the steady state region does not vary much
for a wide range of the oscillation parameters:
for $0.35~{\rm m/s} < V_{max} < 0.75~{\rm m/s}$ (and other parameters fixed),
$\gamma_4$ changes less than 10\%. A similar absence of height dependence
of the velocity distribution function was found in a recent experiment on a
vertically oscillated
quasi-2d granular gas~\cite{vanzon}.

\begin{figure}[t]
{\includegraphics[width=0.8\columnwidth]{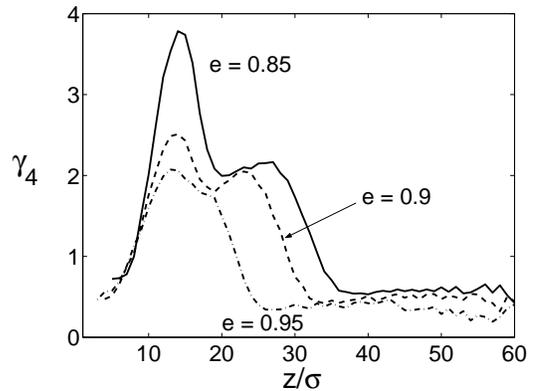}}
\caption{
\label{kurtosis_e}
Kurtosis for three values of $e$, as a function of height
at $ft = -0.2$, where $\mu$ and $\beta_0$ are set to 0.5 and 0.35.
Different forcings are applied for each case to achieve
similar profiles of the hydrodynamic fields:
$V_{max}~(a_{max}) = 0.4$ m/s ($43g$), 0.55 m/s ($60g$), and 0.66 m/s
($72g$) for $e = 0.95$, $e = 0.9$, and $e = 0.85$, respectively.
}
\end{figure}

\begin{figure}[t]
{\includegraphics[width=0.8\columnwidth]{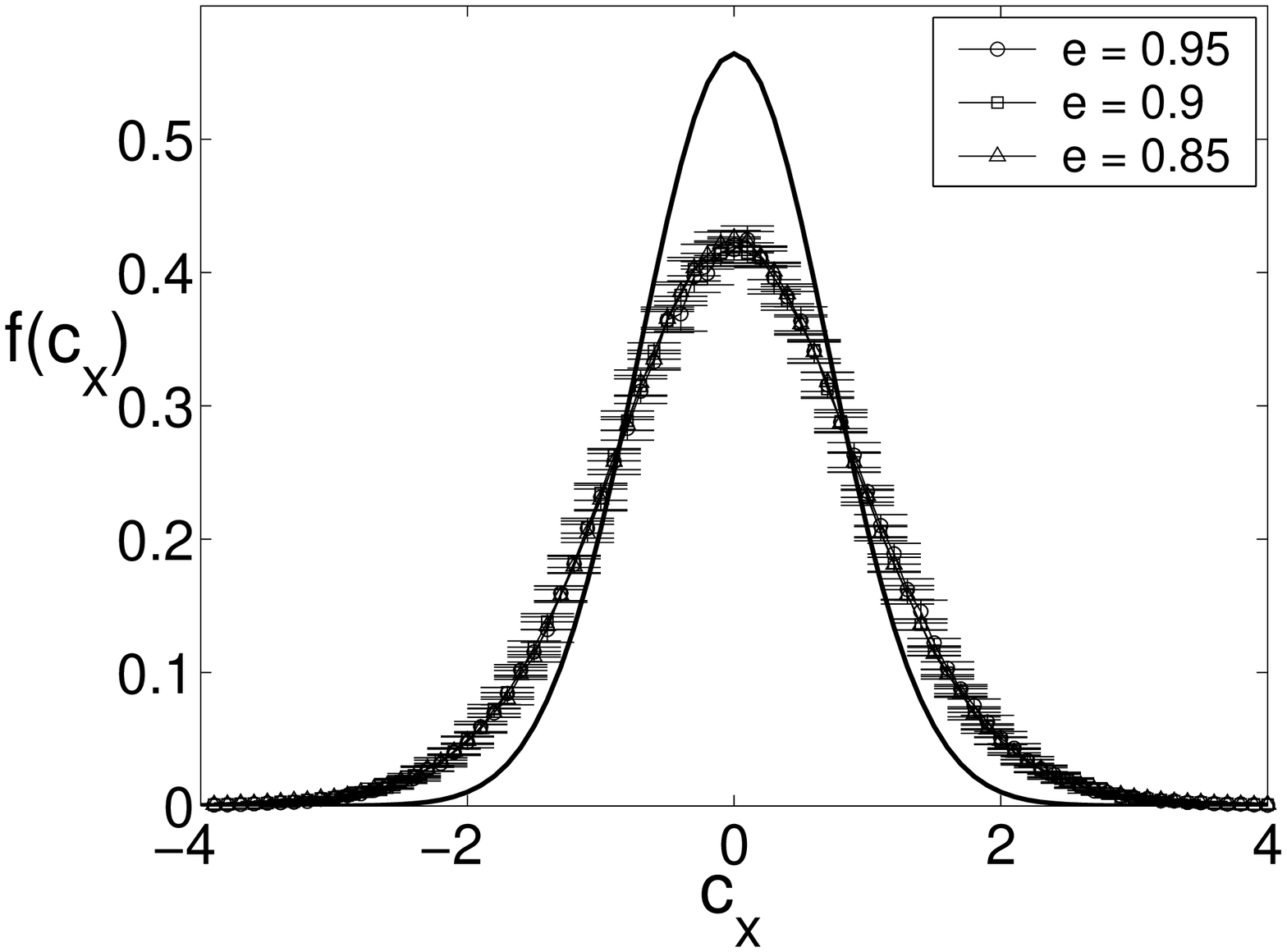}}
{\includegraphics[width=0.8\columnwidth]{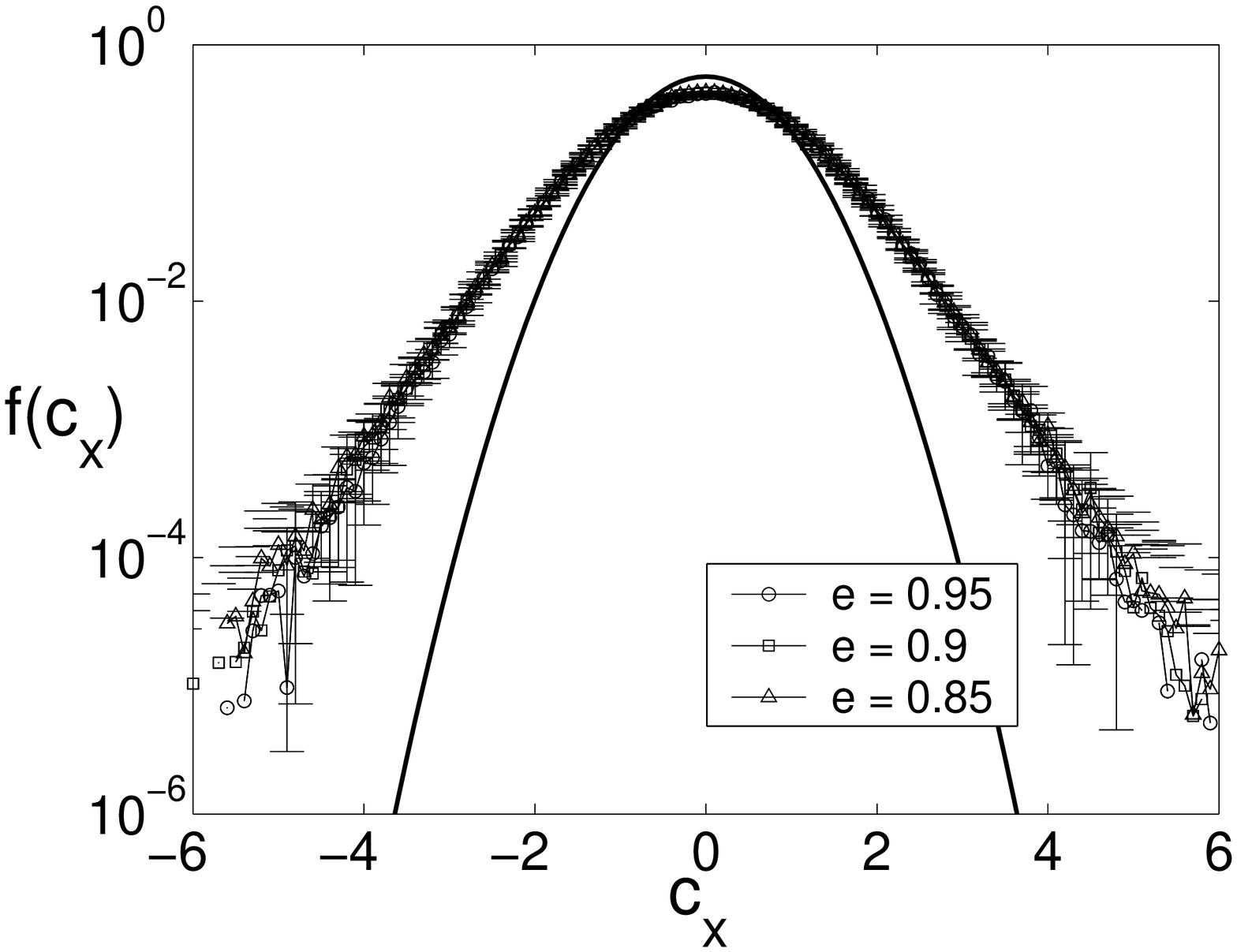}}
\caption{
\label{VDF_e}
The distribution functions of scaled horizontal velocities
$f(c_x)$ for the cases in Fig.~\ref{kurtosis_e},
on linear (top panel) and logarithmic (bottom panel) scales.
Although the kurtosis slightly decreases with $e$,
the difference between the distributions is hardly
distinguishable on both scales.
The ranges for the averaging in height were between $30\sigma$
and $45\sigma$ for $e=0.95$, $40\sigma$ and $55\sigma$ for
$e = 0.9$, and $45\sigma$ and $60\sigma$ for $e = 0.85$;
the averaging was done over relatively similar heights in
the steady state regions (see Fig.~\ref{kurtosis_e}).
The solid line is the MB distribution.
}
\end{figure}

\begin{figure}[t]
{\includegraphics[width=0.8\columnwidth]{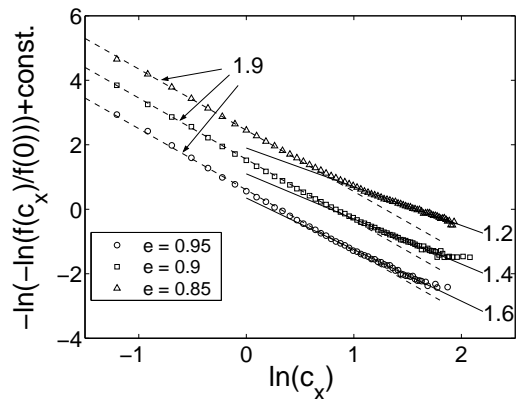}}
\caption{
\label{lnlnYF}
Double logarithm of the distributions of scaled horizontal
velocities for the cases in Fig.~\ref{kurtosis_e}
as a function of the logarithm of $c_x$.
The slope corresponds to the negative exponent, $-{\alpha}$,
of a stretched exponential function $\exp(-c_x^{\alpha})$.
${\alpha}$ is the same for small velocities for three different $e$'s,
however, it depends on $e$ in high energy tails.
Dashed lines correspond to ${\alpha}$ = 1.9, and the solid lines
(from the top) correspond to ${\alpha}$ = 1.2, 1.4, and 1.6,
respectively (indicated by the numbers).
}
\end{figure}

\subsection{Velocity distributions}
We first examine the dependence of the distribution on $e$;
we measure $\gamma_4$ for three different values of $e$,
while $\beta_0$ and $\mu$ are kept at 0.35 and 0.5, respectively.
We find that $\gamma_4$ significantly decreases with increasing $e$
in the oscillatory state, however, it decreases only slightly
in the steady state region (Fig.~\ref{kurtosis_e}).

Now we compare our results with the MB
distribution of variance $1/\sqrt{2}$,

\begin{equation} 
f_{MB}(c_x) = {1 \over \sqrt{\pi}}\exp(-c_x^2).
\end{equation} 
The steady state distributions obtained for the parameters in
Fig.~\ref{kurtosis_e} 
are overpopulated in the high energy tails and underpopulated
at small velocities (Fig.~\ref{VDF_e}), compared to $f_{MB}$.
The differences of the distributions for various $e$'s are hardly
noticeable both on linear and logarithmic scales (Fig.~\ref{VDF_e}), 
but on a double logarithmic scale plot the tails of the distributions
are described by different functions (Fig.~\ref{lnlnYF}).
We investigate the functional form of the distributions
by fitting them (after the normalization by the value at $c_x = 0$)
with a stretched exponential function $\exp(-c_x^{\alpha})$.
We find that the exponent ${\alpha}$ changes from 1.9
(indicated by dashed lines) to some smaller value (solid lines),
depending on $e$, as the velocity increases.
We have not investigated lower values of $e$ to avoid
issues such as cluster formation.

\begin{figure}[b]
{\includegraphics[width=0.8\columnwidth]{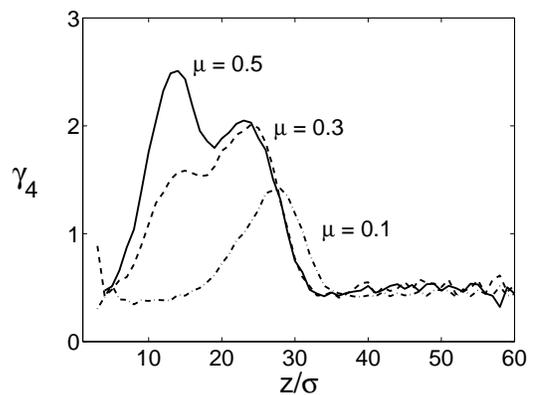}}
\caption{
\label{kurtosis_mu}
Kurtosis of the distributions of scaled horizontal velocities
as a function of height at $ft = -0.2$, for three values
of $\mu$ ($\beta_0$ and $e$ are kept at 0.35 and 0.9, respectively).
In the oscillatory region, $\gamma_4$ increases with $\mu$; however,
$\gamma_4$ is nearly the same within the uncertainty in the steady
state region.
The same forcing ($V_{max} = 0.55$ m/s) is applied to the three cases.
}
\end{figure}

We now keep $e$ and $\beta_0$ at 0.9 and 0.35, respectively,
and change the value of $\mu$.
The profile of $\gamma_4$ in the oscillatory region changes
significantly with $\mu$, however, it is nearly unchanged
in the steady state region (Fig.~\ref{kurtosis_mu}).
The velocity distributions in this region for three different
values of $\mu$ in Fig.~\ref{kurtosis_mu} are also hardly
distinguishable.
We observe that the distribution function depends also on
the density, as in Refs.~\cite{moon01,barrat02,brey03};
however, we do not investigate this dependency systematically.

\subsection{Frictionless inelastic hard sphere model}
In theoretical and numerical studies, granular materials are
often modeled as smooth (frictionless) inelastic hard disks or spheres,
assuming that the friction is a secondary effect that can be
neglected or that both the
inelasticity and the friction can be incorporated together into
a so-called effective coefficient of restitution.
In this Section, we discuss how the velocity distribution changes
when the friction is not included, and we show that the
frictionless model exhibits qualitative differences from
the frictional model.

\begin{figure}[t]
{\includegraphics[width=0.8\columnwidth]{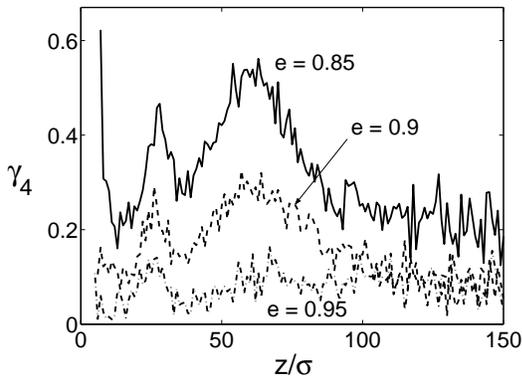}}
\caption{
\label{kurtosis_NF}
Kurtosis of the distributions of scaled horizontal velocities
of frictionless spheres, as a function of height for three
values of $e$ obtained at $ft = -0.2$.
Different forcings (the same as in Fig.~\ref{kurtosis_e})
are applied for each case:
$V_{max}~(a_{max}) = 0.4$ m/s ($43g$), 0.55 m/s ($60g$), and 0.66 m/s
($72g$) for $e = 0.95$, $e = 0.9$, and $e = 0.85$, respectively.
}
\end{figure}

The rotational kinetic energy is two orders of magnitude smaller
than its translational counterpart for the cases studied in this paper.
However, the presence of the friction reduces the expansion of the
granular gas significantly, because the friction reduces the mobility
of the grains and increases the collision frequency~\cite{moon03fric}.
The mean height of frictional inelastic hard spheres exhibits
a different scaling behavior with the plate velocity from
that of frictionless spheres. Only the frictional sphere model
reproduces the experimental observations~\cite{moon03fric,luding95}.

\begin{figure}[b]
{\includegraphics[width=0.8\columnwidth]{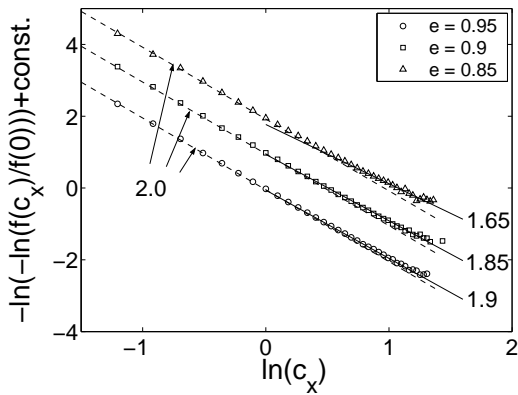}}
\caption{
\label{lnlnNF}
Double logarithm of the rescaled horizontal velocity distribution
functions (normalized by its value at zero) for the cases in
Fig.~\ref{kurtosis_NF}, as a function of the logarithm of $c_x$.
Dashed lines correspond to ${\alpha}$ = 2.0, and the solid lines
(from the top) correspond to ${\alpha}$ = 1.65, 1.85, and 1.9,
respectively.
}
\end{figure}

The $\gamma_4$'s obtained from the simulations of frictionless
particles for the same forcings as in Fig.~\ref{kurtosis_e} are
illustrated in Fig.~\ref{kurtosis_NF}.
In the absence of friction, the layer expands much more,
and the steady state occurs at greater height, $z/\sigma > 100$.
In both the oscillatory and the steady state regions,
values of $\gamma_4$ are smaller than those of frictional spheres
(compare Fig.~\ref{kurtosis_NF} with Figs.~\ref{kurtosis_e}
and ~\ref{kurtosis_mu});
the distribution deviates from $f_{MB}$ only slightly. 
The kurtosis decreases with increasing $e$, and the difference
among the distributions for the three $e$'s are small.
These distributions have four crossovers with $f_{MB}$;
they are overpopulated both at very small and high velocities
and are underpopulated in between, compared to $f_{MB}$.
We fit them with a stretched exponential function, and find
that ${\alpha}$ is 2.0 for small velocities, and
that it depends on $e$ for the high energy tails
(Fig.~\ref{lnlnNF}), as in frictional hard spheres.

\begin{figure}[t]
{\includegraphics[width=0.93\columnwidth]{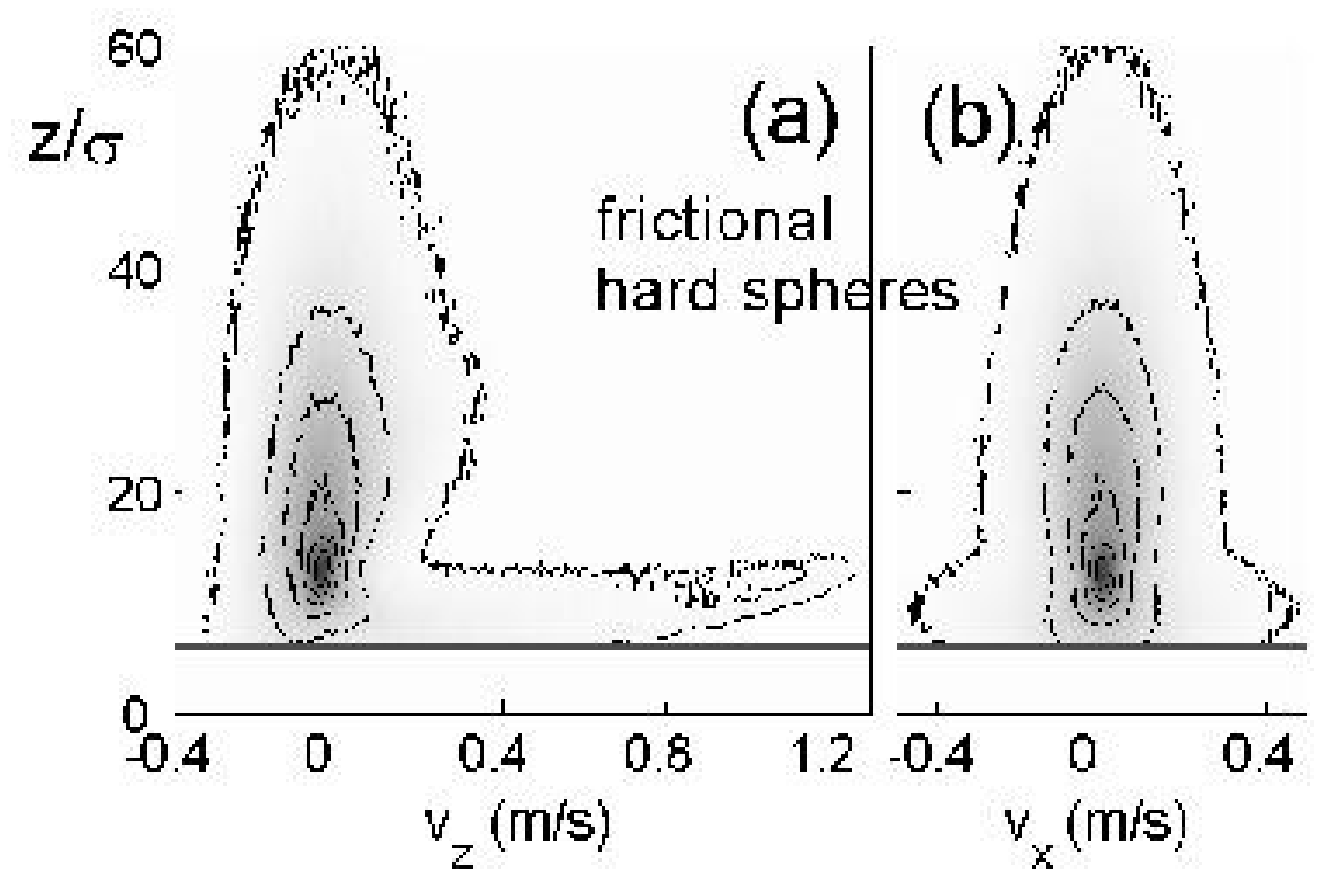}}
{\includegraphics[width=0.93\columnwidth]{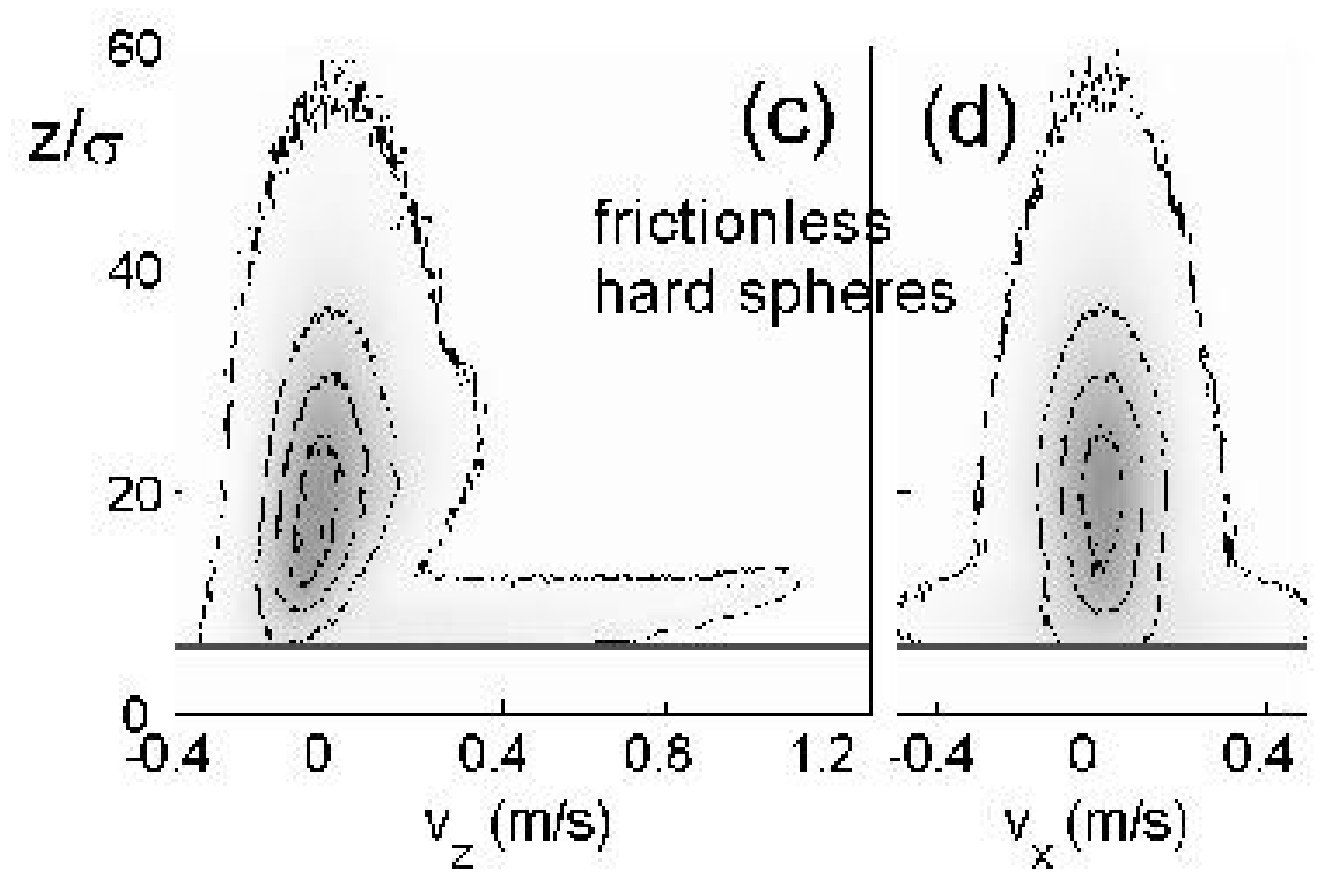}}
\caption{
\label{EffectiveE}
The volume fraction $\nu$ (gray scale and contour lines) at
$ft = 0.25$ as a function of height and $v_z$ [(a) and (c)],
and of height and $v_x$ [(b) and (d)],
obtained from simulations of frictional hard spheres [(a) and (b);
$e = 0.9,~\mu = 0.5$], and of frictionless hard spheres [(c) and (d);
$e = 0.7,~\mu = 0$] at $V_{max} = 0.55$ m/s; $e$ is adjusted in the
frictionless case to obtain comparable overall dissipation and
similar velocity distribution functions in the steady state region.
The frictionless spheres spread more smoothly in height, and
they do not yield the sharp gradient in the density around
$z/\sigma \approx 13$ as in the frictional case, no matter what
the value of $e$ is;
the density profile of the frictionless spheres is qualitatively 
different from that of frictional spheres. Contour lines correspond
to (0.03,0.3,0.6,0.9,1.2,1.5)$\times 10^{-3}$ from outside.
}
\end{figure}

Since the functional form of the distribution depends on $e$,
we can get a similar distribution function for the steady state
by adjusting $e$.
For instance, for $\mu = 0$, $e = 0.7$, and $V_{max} = 0.55$ m/s
(the same forcing as in Fig.~\ref{kurtosis_e}),
we obtain $\gamma_4 \approx 0.5$ for the steady state region;
the steady state distribution is similar to the one in
Fig.~\ref{kurtosis_e} for $e = 0.9$ and $\mu = 0.5$.
However, we find that no single value of $e$ mimics the hydrodynamic
fields or the distribution function of frictional hard spheres,
both in the oscillatory and steady state regions;
the effect of friction cannot be taken over by an adjusted
normal coefficient of restitution.
The difference between the results of the two models is
illustrated in Fig.~\ref{EffectiveE}, where velocities are not
rescaled for better comparison.
The outermost contour lines in both models become similar when
$e$ in the frictionless model is adjusted as a free parameter
[compare Figs.~\ref{EffectiveE}(a) and (c), or (b) and (d)];
however, the overall shape of the density contours cannot be
matched by adjusting only $e$.
Note that the density changes rapidly with height and
$\nu > 1.5 \times 10^{-3}$ at $z/\sigma \approx 10$ near
$v_z \approx v_x \approx 0$ in the frictional model, whereas in the
frictionless model, the particles spread more smoothly in height,
and there is no region for $\nu >1.2 \times 10^{-3}$.

\section{conclusions}
We have studied the horizontal velocity distribution function of
vertically oscillated dilute granular gas, using a molecular dynamics
simulation of frictional, inelastic hard spheres.
The hydrodynamic fields are oscillatory in time near the
oscillating bottom plate due to a shock wave and an expansion wave.
However, the fields are nearly stationary above some height,
thus constituting a granular gas in a nonequilibrium steady state.
The steady state region forms a granular analog of a nearly
collisionless Knudsen gas (Figs.~\ref{CPP} and \ref{mfpKn}).
We find that the dependence of the distribution functions in this granular
Knudsen gas regime on the forcing and material parameters is very
weak, even though the distributions in the collisional bulk at lower heights
depend strongly on the forcing and material parameters
(Figs.~\ref{kurtosis_e} and \ref{kurtosis_mu}).
The behavior of an ordinary Knudsen gas is determined by
boundary conditions~\cite{kogan69}.
Although we do not know whether boundary conditions
or collisions are dominant in determining the behavior of our
granular Knudsen gas, we note that this gas does not
depend much on the properties of its only boundary, which is the
oscillatory region close to the plate.

The functional form of the horizontal velocity distribution
in the steady state region is nearly independent of height,
when velocities are scaled by horizontal temperature (Fig.~\ref{M4}),
even though the hydrodynamic fields continue to change.
The distribution function is broader than the MB distribution,
being underpopulated at small velocities and overpopulated
in the high energy tails (Fig.~\ref{VDF_e}).
We do not observe a universal functional form for the distribution
function (Fig.~\ref{lnlnYF}).
The functional form of the high energy tail changes with the dissipation
parameters ($e$ and $\mu$) and the oscillation parameter ($V_{max}$).
The dependence on $\mu$ in the steady state region is very weak
(Fig.~\ref{kurtosis_mu}).

Our conclusions regarding the absence of a universal distribution
function differ from that of Ref.~\cite{rouyer00}, because:
(1) We studied the local distribution function, while in
Ref.~\cite{rouyer00} the authors obtained the distribution by
averaging over space and time; they assumed that the spatial and
temporal variation was negligible near the center of the oscillating
box, based on their observation of a weak dependence of the density.
We find that the time dependence of the density is weak, but that
of the temperature is strong in the oscillatory state region
(Figs.~\ref{FF} and~\ref{T-INV}).
Note that if a distribution is averaged over different temperatures,
even the MB distribution function leads to a different resultant
distribution function.  (2) Our system is different from that
in Ref.~\cite{rouyer00}: we do not have either air or sidewalls,
and our container is much taller, so that the bottom plate
is the only energy source in our case.
How air and sidewalls affect the dynamics of a granular gas is
yet to be clarified.

We also studied the velocity distributions of frictionless inelastic
hard spheres, and examined the possibility of including frictional
effects using an effective normal coefficient of restitution.
We found that no single effective restitution coefficient could 
describe the frictionless gas at different heights.

Velocities of a granular gas, even in the dilute limit, are strongly
correlated, and the correlations depend on the density
and the coefficient of restitution~\cite{moon01}.
The dependence of the distribution on the density implies that
the single particle distribution of a dilute granular gas cannot
play a role equivalent to that in a dilute ordinary gas;
it is not sufficient to specify statistical properties of the gas.
However, the knowledge of the single particle distribution of
this complex nonequilibrium gas is still of great importance
for the purpose of the first approximation.

The authors thank J. Bougie, D. I. Goldman, E. Rericha,
and J. van Zon for helpful discussions.
This work was supported by DOE Grant DE-FG-0393ER14312 and
Texas Advanced Research Program Grant ARP-055-2001.

\end{document}